\documentclass[10pt,twoside,BCOR7mm,DIV15,headinclude,footexclude,
               cleardoubleempty,idxtotoc]
{scrartcl}

\usepackage{natbib}
\usepackage[font=small,labelfont=bf]{caption}
\usepackage[english]{babel}
\usepackage{graphicx}
\usepackage{hyperref}
\usepackage{scrpage2}
\usepackage{ifthen}
\usepackage{booktabs}
\usepackage{amsmath}
\usepackage{amssymb}
\usepackage{multicol}
\usepackage{float}
\usepackage{hyperref}

\hypersetup{breaklinks=true
,colorlinks=true,linkcolor=black,urlcolor=blue
,citecolor=black}

\addto\captionsenglish{%
}

\makeatletter
\renewcommand{\@biblabel}[1]{}
\renewcommand{\@cite}[2]{%
{#1\ifthenelse{\boolean{@tempswa}}{,#2}{}}}
\makeatother
\ofoot{\thepage}
\ifoot{}

\setcapindent{0em}
\setheadsepline{1pt}

\setkomafont{pagehead}{\normalfont\sffamily}
\setkomafont{pagenumber}{\normalfont\rmfamily}

\makeatletter
\newcommand{\listofcontributions}{\@starttoc{con}}

\newcommand{\l@contribution} {\@dottedtocline{1}{1.5em}{2.3em}}
\makeatother

\newenvironment{contribution}{
\setcounter{section}{0}
\setcounter{figure}{0}
\setcounter{table}{0}
}{
\newpage
\lehead{}
\rohead{}
}

\def\aj{AJ}%
\def\araa{ARA\&A}%
\def\apj{ApJ}%
\def\aap{A\&A}%
\def\mnras{MNRAS}%
\def\nar{New A Rev.}%
\def\ssr{Space~Sci.~Rev.}%
\begin{document}

\setlength{\baselineskip}{2.5ex}

\begin{contribution}
% EXAMPLE AND TEMPLATE FILE FOR PROCEEDINGS OF THE WOLF-RAYET WORKSHOP.
% PLEASE REPLACE THE TEMPLATE TEXT BY YOUR OWN ARTICLE.
% NOTE THAT YOU MUST NOT PROCESS THIS FILE, BUT THE MASTER FILE:
% latex masterfile; dvips masterfile

% RUNNING AUTHOR: PUT AUTHOR NAMED HERE
\lehead{P.A. Crowther}

% RUNNING TITLE; SHORTEN THE TITLE IF NECESSARY
% IN CASE OF A ONE-PAGE CONTRIBUTION (POSTER),
% SQUEEZE AUTHORS AND TITLE IN THIS LINE (Author: Title ...)
\rohead{WR content of the Milky Way}

\begin{center}
% FULL TITLE HEADING
{\LARGE \bf Wolf-Rayet content of the Milky Way}\\
\medskip

% AUTHORS LIST
{\it\bf Paul A. Crowther}\\

% AFFILIATIONS
{\it Dept of Physics \& Astronomy, University of Sheffield, Sheffield S3 7RH, UK}

% ABSTRACT
\begin{abstract}
An overview of the known Wolf-Rayet (WR) population of the Milky Way is presented, including
a brief overview of historical catalogues and recent advances based on infrared photometric and spectroscopic 
observations resulting in the current census of 642 (v1.13 online catalogue). 
The observed distribution of WR stars is considered with respect to known star clusters, given that $\leq$20\% of
WR stars in the disk are located in clusters. WN stars outnumber WC stars at all galactocentric radii, while early-type
WC stars are strongly biased against the inner Milky Way. Finally, recent estimates of the global WR population in the 
Milky Way are reassessed, with 1,200$\pm$100 estimated, such that the current census may be 50\% complete.
A characteristic WR lifetime of 0.25 Myr is inferred for an initial mass threshold of 25 $M_{\odot}$.
\end{abstract}
\end{center}

% TEXT OF THE PAPER, TWO-COLUMN STYLE
\begin{multicols}{2}

\section{Historical overview and current census}

Wolf-Rayet (WR) stars are the evolved descendents of massive stars, with two main flavours: Helium-rich WN stars displaying the products of core H burning and Carbon-rich WC stars displaying the products of core He burning \citep{2007ARA&A..45..177C}. An overview of the number of  WR stars in the Milky Way is provided. The Galactic census tripled between the 1st \citep{1894AstAp..13..448C} and the 6th catalogues (van der Hucht 1981), a century later (Table~\ref{catalogues}), and subsequently doubled by the Annex to the 7th catalogue \citep{2006A&A...458..453V} owing to deep narrow-band optical surveys \citep{1999AJ....118..390S}, the discovery of rich star clusters \citep{2006MNRAS.372.1407C} and infrared surveys of the Galactic Centre region \citep{2002ApJ...581..258F}.
Over the past decade, the number has doubled again, primarily through surveys arising from the advent of large format infrared detectors, with the
census totalling 642\footnote{v1.13 of online WR catalogue http://pacrowther.staff.shef.ac.uk/WRcat/}, 
comprising 357 WN stars, 8 WN/C stars, 273 WC stars and 4 WO stars. 

\begin{table}[H] 
\begin{center} 
\captionabove{Historical catalogues of Galactic WR stars} 
\label{catalogues}
\begin{tabular}{c@{\hspace{2mm}}l@{\hspace{2mm}}c}
\toprule
Catalogue  & Reference  & Number \\
\midrule 
I & \citet{1894AstAp..13..448C} & 55 \\
II & \citet{1912AnHar..56..165F}  & 108 \\
III & \citet{1930HarMo...3....1P} & 92 \\
IV & \citet{1962AJ.....67...79R} & 123 \\
V & \citet{1968MNRAS.138..109S} & 127 \\
VI & \citet{1981SSRv...28..227V} & 157 \\
VII &  \citet{2001NewAR..45..135V} & 227 \\
VII% a & \citet{2006A&A...458..453V} & 298 \\
      & Current census v1.13 & 642$^{1}$ \\ 
\bottomrule
\end{tabular}
\end{center}
\end{table}

Two complementary initiatives have dominated this improvement since the annex to the 7th catalogue \citep{2006A&A...458..453V}. Mike Shara and collaborators have employed narrow-band imaging of the Galactic plane at 2-2.3$\mu$m to identify 150+ WR stars from follow-up spectroscopy \citep{2012AJ....143..149S}. Since this approach relies upon the contrast between emission lines and the adjacent continuum, systems which are dominated by hot dust or a companion in the K-band will not be identified. A second method exploiting the unusual near- to mid-IR broad-band colours of WR stars has been
exploited by Schuyler Van Dyk, Pat Morris and collaborators has identified 100+ WR stars 
\citep{2007MNRAS.376..248H, 2011AJ....142...40M}. This approach has successfully identified dusty WR stars, but would also prove problematic for systems whose IR energy distribution is dominated by a companion star.

\begin{figure}[H]
\begin{center}
\includegraphics[width=\columnwidth]{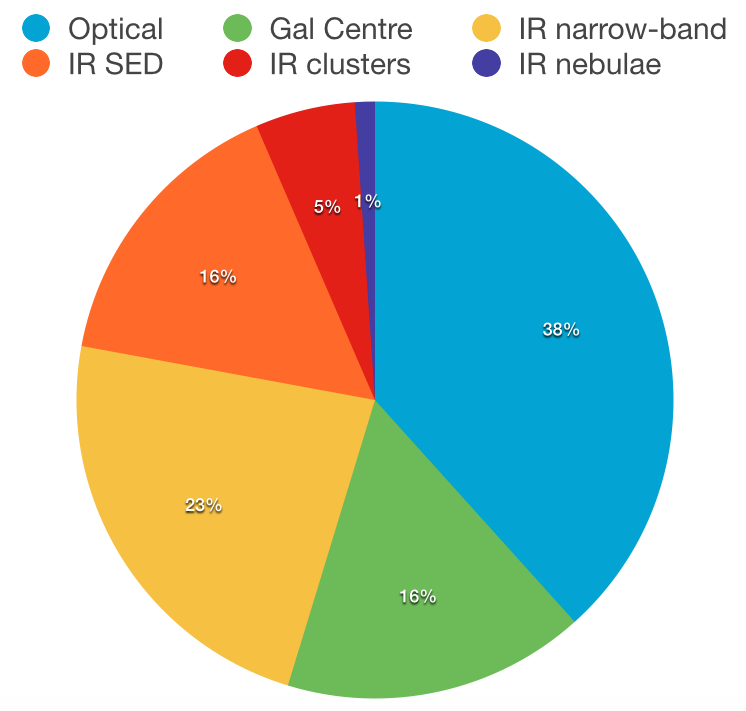}
\caption{Pie chart illustrating breakdown of discovery technique for WR stars in current census (v1.13 of online catalogue)}
\label{wr_surveys}
\end{center}
\end{figure}

In addition, WR stars have been discovered from follow-up spectroscopy of members in star clusters newly identified from IR surveys (e.g. Spitzer/GLIMPSE,
VISTA/VVV) such as \citet{2012A&A...545A..54C}. Other WR stars have been newly identified from follow-up spectroscopy
of the ionizing stars in dusty mid-IR nebulae \citep{2010AJ....139.2330W}. A chart illustrating the fraction of WR stars discovered by different techniques is illustrated in Fig.~\ref{wr_surveys}.

The binary frequency of O stars in young Galactic clusters is believed to be high: $\sim$71\% will interact with a 
companion during their lifetime according to \citet{2012Sci...337..444S}. Since WR stars are the progeny of O stars, a
high WR binary frequency will also be anticipated. The majority of searches for binarity in WR stars have been
conducted via blue visual spectroscopic monitoring.  Amongst the 25\% of WR stars which are visually brighter
than V=15 mag, 44 are close binaries (either SB1 or SB2), plus another 10 are dusty WC stars. It is now widely accepted that dusty WC stars are universally binary systems involving a WC star and an O star companion. 
Therefore,  a strict lower limit to the binary frequency is 34\% (= 54/160 stars). For comparison, a WR binary frequency of 38\% (86/227) was
estimated in the 7th WR catalogue \citep{2001NewAR..45..135V}, including long period systems (P $>$ 1000 days) and X-ray bright systems - by way of example the X-ray bright system WR25 was subsequently identified as an SB1 by \citet{2006A&A...460..777G}. These statistics undoubtedly represent lower limits to the WR binary frequency, although one would expect a lower binary frequency than for O stars since some WR stars will be merger products and others will become single following the (dynamical or post-supernova) disruption of initial binary systems.

\section{WR stars and star clusters}

If we adopt a lower mass limit to the production of WR stars of $\sim$25 $M_{\odot}$, and if massive stars only form in clusters, whose upper mass limit follows the relationship of \citet{2010MNRAS.401..275W}, one would expect 10$^{3} M_{\odot}$ clusters to host the majority of WR stars. In reality only 27\% of WR stars in the Milky Way are in known star clusters \citep{2015MNRAS.447.2322R}. Of these, the majority of known WR stars in the Central Molecular Zone (CMZ) do lie within one of the three massive star clusters, the Arches, Quintuplet and the Galactic Centre cluster itself. Excluding the CMZ, only 18\% of the WR stars located in the disk lie within star clusters. This is illustrated in Fig.~\ref{clusters}. 

Consequently, either the lower
mass limit to the production of WR stars must be significantly lower (making birth clusters harder to identifity), WR stars do not preferentially form in dense star clusters, or the majority of WR stars are no longer associated with their birth cluster. 

OB stars are observed in a range of environments - low density star forming regions (e.g. $\rho$ Oph), intermediate density OB associations such as Cyg OB2, and dense clusters such as NGC~3603. Indeed, the fraction of stars (of all masses) located in dense (Orion Nebula-like) clusters in the Solar Neighbourhood is $<$26\% \citep{2010MNRAS.409L..54B}.  \citet{2014MNRAS.438..639W} have established that massive stars in Cyg OB2 did not form in close proximity, nor in regions of higher density, so can form in regions of relatively low density.  

Therefore, it is likely that typical massive stars arise from relatively loose OB associations, with only a small fraction born in dense star clusters, so the rarity of WR stars in star clusters naturally follows without resorting to a low threshold mass to the formation of WR stars or a high ejection frequency. This has relevance to the discussion of \citet{2015MNRAS.447..598S} regarding the relative masses of O, WR stars and Luminous Blue Variable (LBV) in the Large Magellanic Cloud from their spatial locations. One would expect relatively few O stars in close proximity to most WR stars/LBVs if the majority of massive stars originate in intermediate density environments, especially since most WR stars will arise from relatively modest $\sim 25 M_{\odot}$ progenitors.

\section{Distribution of WR stars}

Since the distance of the majority of Galactic WR stars is not well established, \citet{2015MNRAS.447.2322R} have investigated the near-IR absolute magnitudes of different WR subtypes based on 108 stars whose distances have been estimated, either from cluster/association membership or other techniques. The spatial distribution of these, plus 246 field WR stars for which reliable spectral types have been obtained, are presented in Fig.~\ref{wr_distribution}. 

Field WR populations include binary systems, providing the WR to OB light ratio can be estimated in the K-band or the WR star is thought to dominate the near-IR continuum flux. From a census of local ($<$3 kpc) WR systems, the WR star dominates the K-band flux in 82\% of cases. In addition, dusty WC stars are omitted from this distribution because their K-band flux is typically dominated by hot dust rather than the WC continuum. For reference, dusty WC stars comprise $\sim$15\% of the local ($<$3 kpc) WR population.

Generally, the near-IR classification of WR stars is rather coarser than at visible wavelengths. The majority of near-IR studies have followed the scheme of \citet{2006MNRAS.372.1407C} who utilised IRTF/SpeX 1--5$\mu$m observations of $\sim$30 optically classified WR stars. In general robust near-IR classification requires observations at J, H and K, although the actual number of WN/C
transition stars will be underestimated since no near-IR schemes have so far been implemented (efforts are currently
underway to remedy this deficiency).

\begin{figure}[H]
\begin{center}
\includegraphics[width=0.89\columnwidth]{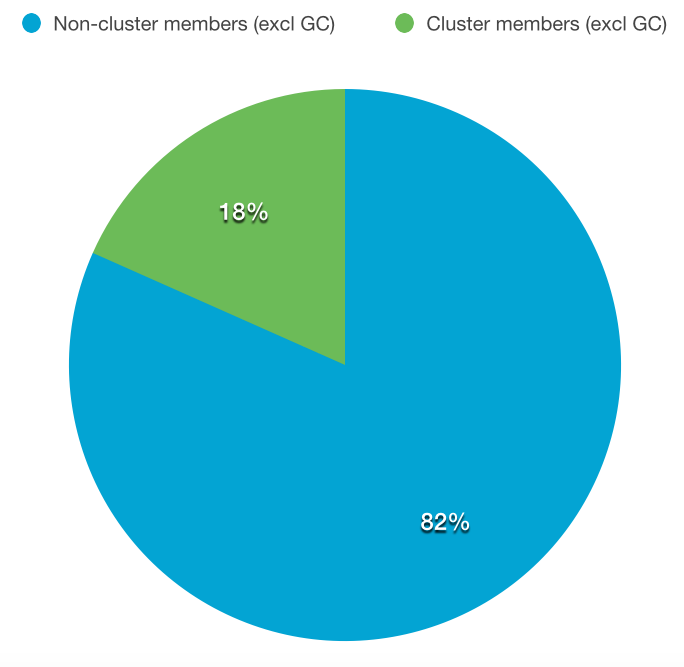}
\includegraphics[width=0.82\columnwidth]{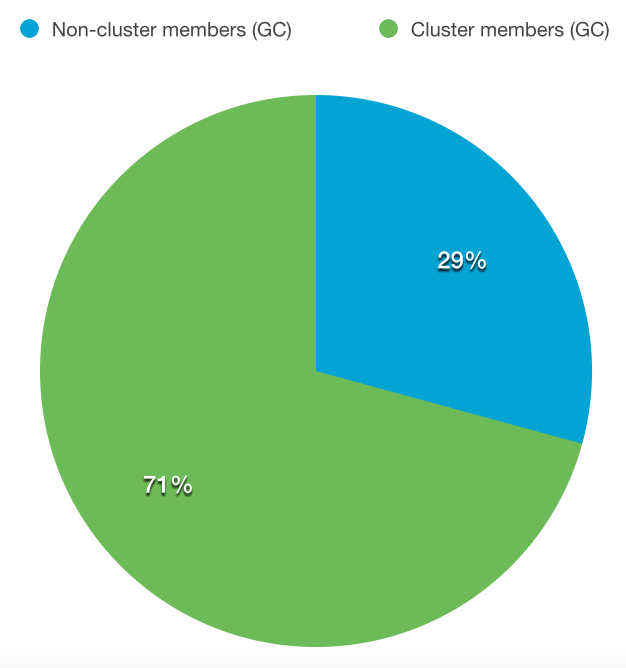}
\includegraphics[width=0.77\columnwidth]{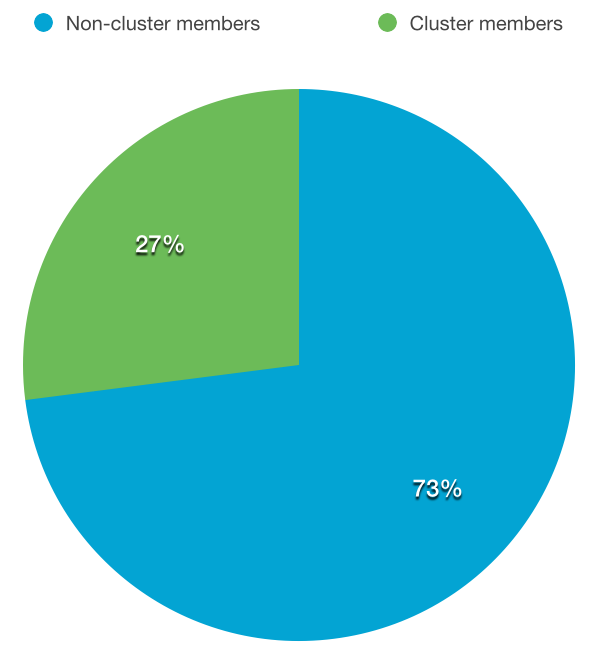}
\caption{Pie charts illustrating the fraction of Wolf-Rayet stars (v1.13 of catalogue) which are cluster 
members (green) or in the field (blue) for the Galactic disk (top), Galactic Centre region (middle) and combined pop (bottom).}
\label{clusters} 
\end{center}
\end{figure}

The observed ratios of WR subtypes in the inner Galaxy at galactocentric distances of below 6 kpc ($\log$ O/H+12 $\sim$ 8.85), mid disk from 6--9 kpc ($\log$ O/H+12 $\sim$ 8.7) and outer disk beyond 9 kpc ($\log$ O/H + 12 $\sim$ 8.55) is presented in Table~\ref{ratios}. Overall there is little variation between these regions with the exception of early to late WC stars, with the latter dominating in the inner disk owing to the metallicity dependence of WC classification diagnostics  \citep{2002A&A...392..653C}. Incorporating results from the Magellanic Clouds, the observed 
WC to WN ratio is presented in Fig.~\ref{wcwn}, together with a variety of evolutionary predictions from single and binary models (see caption).

\begin{figure}[H]
\begin{center}
\includegraphics
  [width=\columnwidth]
  {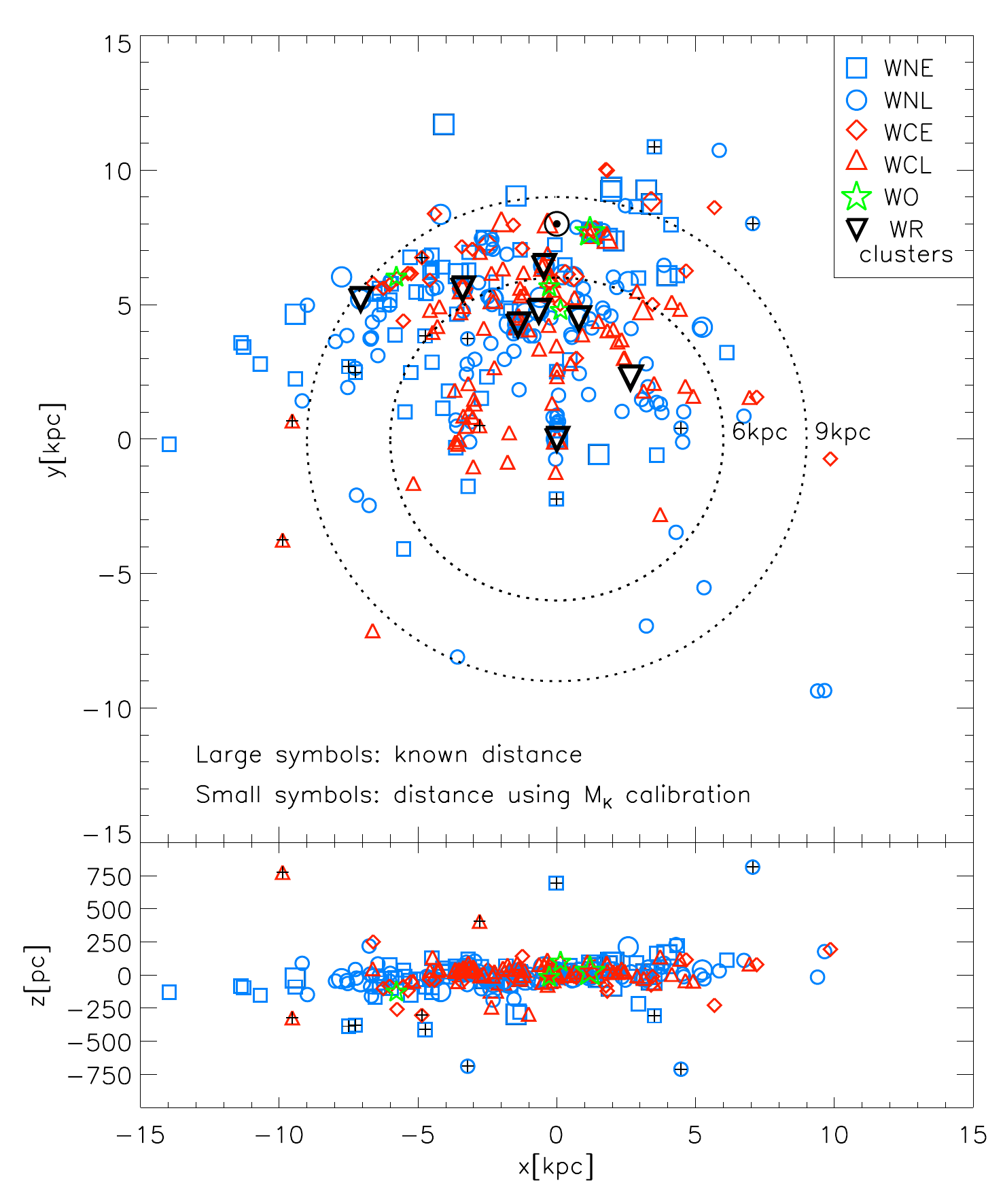}
\caption{Distribution of 354 WR stars in the Galactic disk from \citet{2015MNRAS.447.2322R} }
\label{wr_distribution} 
\end{center}
\end{figure}

\section{Global WR content}

Various estimates of the global WR population in the Milky Way have been made, ranging from 1,200 \citep{1982A&A...114..409M} to 6,500 \citep{2001NewAR..45..135V}. We have constructed a toy model of the 
WR population in the Milky Way in an azimuthally symmetric disk following the radial HII distribution, atomic/molecular dust distribution and the observed WN/WC distribution \citep{2015MNRAS.447.2322R}.

\begin{table*}[!t] 
\begin{center} 
\captionabove{WR subtype distribution in the Milky Way \citep{2015MNRAS.447.2322R} for three galacto-centric
distances ($R_{\rm GC})$.} 
\label{ratios}
\begin{tabular}{crccc}
\toprule
Region & N(WR) & N(WC)/N(WN) & N(WC+d)/N(WN) & N(WCE)/N(WCL) \\
\midrule 
Inner ($R_{\rm GC} < 6$ kpc) & 187 & 0.51 &  0.69 & 0.05 \\
Mid (6 $\leq R_{\rm GC} \leq$ 9 kpc)&   132 & 0.53 & 0.73 & 1.0 \\
Outer ($R_{\rm GC} > $ 9 kpc) & 35 & 0.40 & 0.57 & 1.5 \\
\bottomrule
\end{tabular}
\end{center}
\end{table*}

If we assume that the observed WR distribution is complete to K=8 mag (though see below), $\sim$550 WR dominated stars in the Galactic disk reproduce the observed WR distribution. In addition, over 100 WR  stars are currently known with $-2.5^{\circ} \leq l \leq +3.5^{\circ}$ so the total CMZ population may be as large as 250.  Such a large population is surprisin given that the CMZ accounts for perhaps 5\% of the entire Milky Way star formation rate \citep{2013MNRAS.429..987L}, but the presence of three very massive clusters in the Galactic Centre region suggests WR stars may be disproportionately represented.  

\begin{figure}[H]
\begin{center}
\includegraphics[width=\columnwidth]{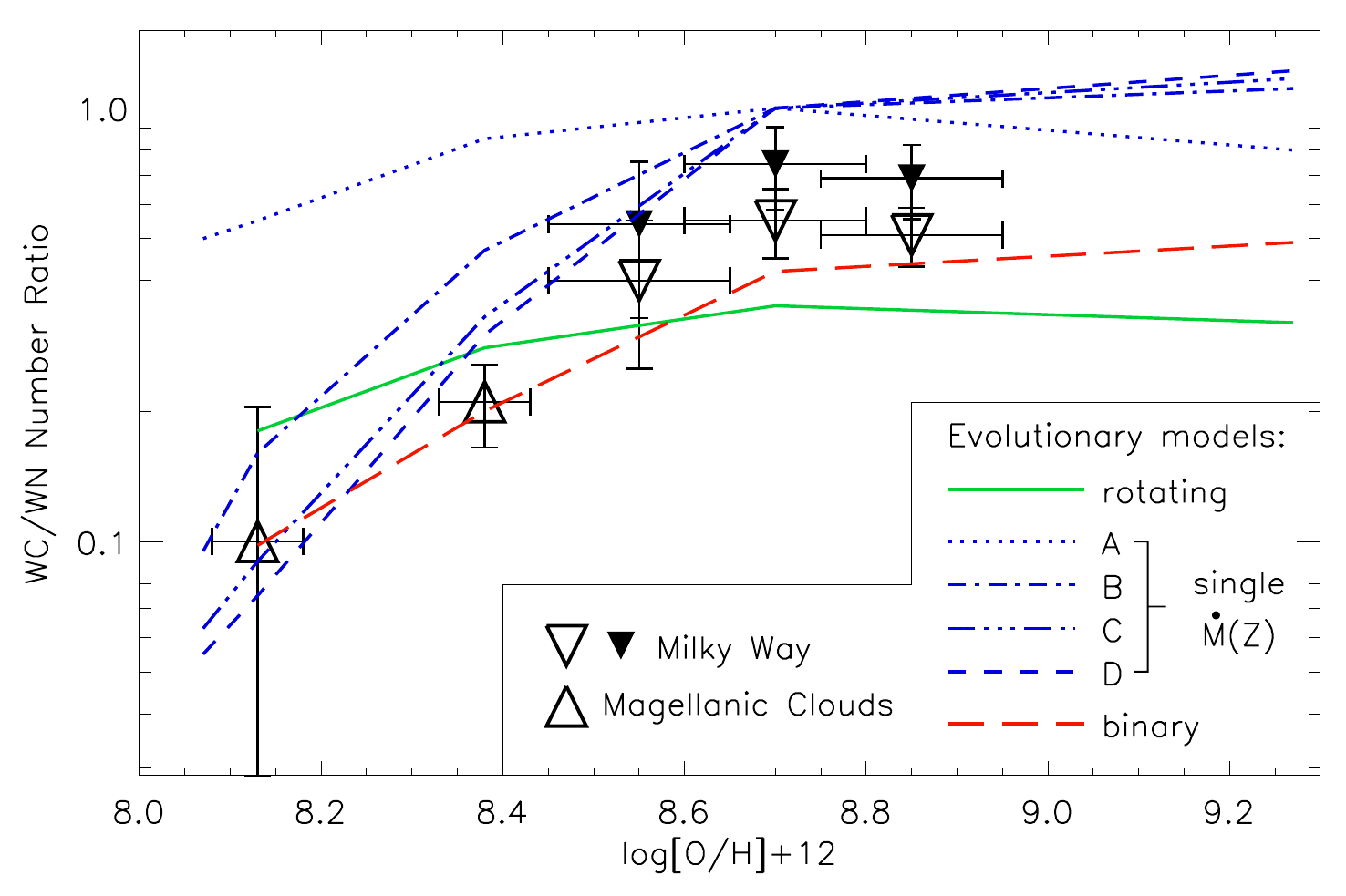}
\caption{Ratio of WC to WN stars in the Milky Way, for which dusty WC stars have been omitted (inverted open 
triangles) or included (inverted filled triangles), plus the Magellanic Clouds (open triangles) from \citet{2015MNRAS.447.2322R}. Predictions from rotating single star models \citep[green]{2005A&A...429..581M}, binary 
models \citep[red]{2008MNRAS.384.1109E} and non-rotating single star models for various mass-loss metallicity dependencies \citep[blue]{2006A&A...452..295E} are also indicated.}
\label{wcwn} 
\end{center}
\end{figure}

A total of 550 + 250 = 800
stars would represent only 82\% of the non-dusty WR population since 18\% of systems are expected to be dominated by the WR companion in the K-band, so the non-dusty WR population inferred is $\sim$950. Finally, including dusty WC stars, which comprise 15\% (150) of the whole population, we estimate N(WR) = 1,100, providing N(WCd)/N(WC) is uniform across all galactocentric radii. A histogram of the predicted JHK$_{s}$ magnitudes for this predicted
WR population is presented in Fig.~\ref{histogram}.

\begin{figure}[H]
\begin{center}
\includegraphics[width=\columnwidth]{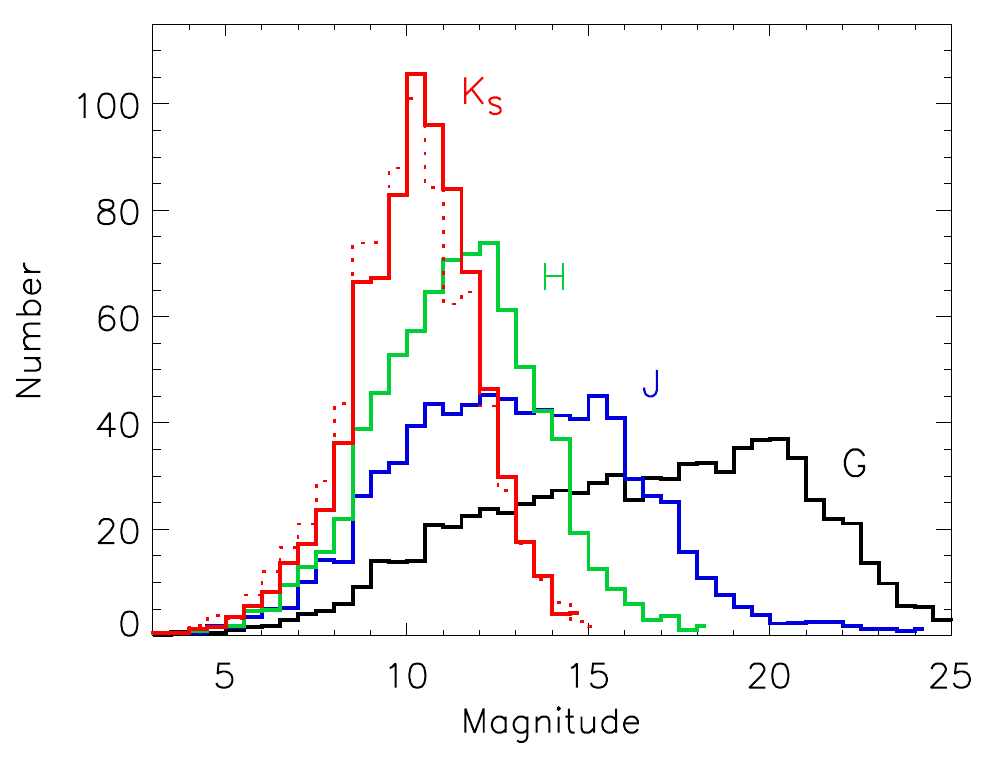}
\caption{Histogram of JHK$_{s}$ and GAIA G-band magnitudes predicted from the model WR population by \citet{2015MNRAS.447.2322R}.Thick lines represent populations consisting of WN and non-dusty WC stars, while the dotted-line illustrates a model population in which 28\% of WC stars are dust-forming with $M_{\rm K} = -6.95$ mag.}
\label{histogram} 
\end{center}
\end{figure}

Based on this global WR population of 1,100, if the lower mass threshold to WR  stars is $25 M_{\odot}$, an average lifetime for the WR phase of 0.25 Myr is implied by the current Milky Way star formation rate of 1.9 $M_{\odot}$ yr$^{-1}$  \citep{2011AJ....142..197C} and a standard Kroupa Initial Mass Function. This characteristic lifetime is in fair agrement with non-rotation evolutionary models, but is somewhat shorter than rotating model predictions \citep{2012A&A...542A..29G}.

If the observed WR distribution is not complete to K=8 mag, the acual WR population will increase. By way of example, the two most recent additions to the on-line catalogue (v1.13) involve WR111-13\footnote{All WR stars discovered between the 6th catalogue and the Annex to the 7th catalogue follow the WRXXXa, WRXXXb nomenclature, whereas more recent additions utilise WRXXX-1, -2}, a WN6 star from \citet{2015ApJ...805..110M} with K=8.02 mag and WR111-14, a WN7--8 star from \citet{2015MNRAS.452..884N} with K=7.7 mag. If the WR-dominated disk population were 650 instead of 550, a global population of 1,300 would be inferred, so the current census may be 50\% complete.

\section{Summary}

A brief overview of the Wolf-Rayet population in the Milky Way is presented. The current census of 642 (v1.13 of online catalogue) represents a doubling of the population since the Annex to the 7th WR catalogue only a decade ago, due to systematic searches from narrow-band surveys at 2$\mu$m and broad-band near to mid-IR surveys. 

A strict lower limit to the binary fraction is 34\% (SB1--2 systems plus dusty WC stars) for V$<$ 15 mag, but the actual fraction will oudoubedly be significantly higher. Spectroscopic monitoring of WR stars in the near-IR is now feasible owing to multi-object IR integral field units (e.g. KMOS at ESO's VLT). 

Excluding the Galactic Centre region, $\leq$20\% of WR stars are located in star clusters, adding weight to the formation of most massive stars in OB associations, i.e. away from dense clusters. The current WR census may be 50\% complete, in which case the average duration of the WR phase is 0.25 Myr for a lower mass limit of 25 $M_{\odot}$ to the formation of WR stars. 

The majority of systems hosting WR stars are dominated by the He-rich component in the K-band, so these are fundamentally high luminosity (high mass) stars. In contrast, typical progenitors of stripped envelope core-collapse supernovae are relatively low mass stars, having transferred the majority of their H-rich envelope to a close companion. Such systems would not be recognised as conventional WR stars since the He-rich mass donor would be masked by the H-rich mass gainer at optical and near-IR wavelengths. 

Finally, the online WR catalogue is maintained in Sheffield on a best efforts basis  so authors are encouraged to provide Paul Crowther with preprints as soon as they are accepted for publication.

\end{multicols}
\end{contribution}

%%-------------------------------------------------------

\end{document}